\documentclass[a4paper, 12pt]{article}
\begin{document}
\begin{center}
{\Large\textbf{On the Absence of Cross--Confinement for 
\vskip10pt
Dynamically Generated Multi--Chern--Simons 
\vskip16pt
Theories}}\\
\vspace {8mm}
\renewcommand{\thefootnote}{$\ast$}
{\large Emil M. Prodanov\footnote {Supported by FORBAIRT scientific research
program and Trinity College, Dublin.} and Siddhartha Sen} \\
\vspace {2mm}
{\it School of Mathematics, Trinity College, Dublin 2, Ireland, \\
e-mail: \hskip 2pt prodanov@maths.tcd.ie, \hskip 2pt sen@maths.tcd.ie}
\end{center}
\vspace{4mm}
\begin{abstract}
We show that when the induced parity breaking part of the effective action for
the low--momentum region of $U(1) \times \ldots \times U(1)$ Maxwell gauge 
field theory with massive fermions in 3 dimensions is coupled to a 
$\phi^{\scriptscriptstyle 4}$ scalar field theory, it is not possible to 
eliminate the screening of the long-range Coulomb interactions and get 
external charges confined in the broken Higgs phase. This result is valid for
non-zero temperature as well.
\end{abstract}
\scriptsize
{\bf PACS numbers}: 11.10.Jj, 11.10.Kk, 11.10.Wx, 11.15.Ex \\
{\bf Keywords}: Finite Teperature Field Theory, Spontaneous Symmetry Breaking,
Chern--Simons, Screening, Confinement. 
\normalsize
\newpage
\def\be{\begin{equation}}
\def\ee{\end{equation}}
\def\s{\scriptscriptstyle}
\def\ba{\begin{eqnarray}}
\def\ea{\end{eqnarray}}
\def\sl{\hskip3pt \slash \mkern-13mu}
\section{Introduction}
In a recent paper Cornalba et al.~\cite{wil} have proposed a novel topological
way of confining charged particles. The method uses the special properties
of $U(1) \times U(1)$ Chern--Simons gauge theory, interacting with external
sources in two spatial dimensions, with a scalar Higgs field providing 
condensates. The idea of the approach is to note that, when charge/flux 
constraints of a certain type are not satisfied, the fall off of the Higgs 
fields at infinity will not be fast enough and will lead to configurations 
with infinite energy; hence, such configurations are confined. The analysis is
based on number--theoretic properties of the couplings and charges and shows 
the intriguing possibility for confinement even for integral charge particles.
The confinement mechanism is topological in origin.

A Chern--Simons term of the form considered in~\cite{wil} can be dynamically 
generated as the parity--breaking part of the low--momentum region of the 
effective action of a three--dimensional $U(1) \times \ldots \times U(1)$ 
Maxwell gauge filed theory with fermions, after integrating out the fermionic 
degrees of freedom~\cite{schap},~\cite{deser}. Indeed, we carry out this 
procedure for the system at non-zero temperature. The effective action, 
obtained by us, following the approach of~\cite{schap} has the correct 
temperature dependence for the multiple $U(1)$ Chern--Simons term and yields 
in its zero-temperature limit a multiple Chern--Simons term of the form 
considered in~\cite{wil},~\cite{prop1},~\cite{prop2}.

Such multiple $U(1)$ gauge theories have been considered before, for example: 
in the study of spontaneously broken abelian Chern--Simons 
theories~\cite{prop1},~\cite{prop2}; in the study of two--dimensional 
superconductivity without parity violation~\cite{dm}. 

Our original motivation was to investigate if the mechanism for 
cross--confinement, proposed in~\cite{wil}, continues to hold for the system
with a temperature slightly deviated from zero and if confinement is lost for
high temperature with the system still in the Higgs phase.

Surprisingly, with this dynamically generated parity--breaking term, the 
arguments of Cornalba et al.~\cite{wil} do not hold, namely, the proposed 
scheme of confinement is not possible. This result is valid, as we show, 
for zero and non--zero temperatures. In this model it is not possible to
eliminate the screening of the long--range Coulomb interactions. We claim that,
if confinement occurs, it happens when the broken 
$U(1) \times\ldots\times U(1)$ gauge symmetry is restored in at least one of 
the directions of the gauge group.

By the standard Higgs mechanism, the gauge group is spontaneously broken down 
to a product of the cyclic groups
$Z_{\scriptscriptstyle 1} \times\ldots\times Z_{\scriptscriptstyle N}$.
This residual symmetry represents the non-trivial holonomy of the Goldstone
boson. The photon fields $A_{\mu}^{\s (i)}, \hskip 4pt i=1, ... ,N$ now 
acquire masses by their coupling to the Goldstone bosons. In the broken Higgs 
phase the Higgs currents are proportional in magnitude to the massive vector 
fields and screen the Coulomb interaction and we are left with purely quantum 
Aharonov--Bohm interactions~\cite{prop1},~\cite{prop2}. 
In this phase, at temperature well below the critical, all conserved charges 
can reside in the zero--momentum mode due to the bosonic character of the 
particles. When the temperature increases, some of the charges get excited out 
of the condensate and at sufficiently high temperature the condensate becomes 
thermally disordered and the symmetry is restored. When this happens the 
charges introduced by the matter currents will not be screened and the energy 
of the Coulomb field will logarithmically diverge with distance (in two 
spatial dimensions) and this will lead to confinement. 

\section{The Model}
We will determine first the parity--breaking part of the effective 
action for $U(1) \times \ldots \times U(1)$ Maxwell gauge field theory coupled
to massive fermions and $\phi^{\s 4}$ scalar field theory in 3 dimensions at 
finite temperature. Contact with the multiple Chern--Simons term, considered 
by~\cite{wil},~\cite{prop1},~\cite{prop2} is made by taking the 
zero-temperature limit. The effective action for the low--momentum region of 
the theory is:
\ba
e^{-\Gamma(A^{\s (k)}, M_{\s k})} = 
\int\!
\prod\limits_{\s k=1}^{\s N}\!
\mathcal{D}\psi_{\s k}\mathcal{D}\bar{\psi_{\s k}}\mathcal{D}\phi \ 
&& \mkern-45mu
exp \Biggl\{- \int\limits_{\s 0}^{\s \beta} \! d \tau \!\!\int \!d^{\s 2} x 
\biggl\{\sum\limits_{\s k=1}^{\s N} 
\Bigl(\bar{\psi}_{\s k} \sl D^{\s f}_{\s k} \psi_{\s k} + 
j^{\s (k)}  A^{\s (k)}\Bigr) +
\nonumber \\ 
&+& \mkern-10mu
(D_{\s \alpha} \phi)(D^{\s \alpha} \phi)^{*} - m^{\s 2} \phi \phi^{*} 
-  \lambda (\phi \phi^{*})^{\s 2} 
\biggr\}\Biggr\} 
\ea
where $\sl D^{\s f}_{\s k} = \hskip4pt \slash \mkern-9mu \partial + 
i Q_{\s k j} \sl A^{\s (j)} + M_{\s k}$ are the fermionic covariant derivatives
with $Q_{\s kj}, \hskip4pt k,j = 1, \ldots, N$ being the matrix of the 
fermionic charges with respect to the $N$ gauge groups, $D_{\s \alpha} = 
\partial_{\s \alpha}+i q_{\s k} A^{\s (k)}_{\s \alpha}, k=1, \ldots, N$ 
are the covariant derivatives for the scalar field with $q_{\s i}$ being the 
charge of the scalar field with respect to the $i^{\s th}$ gauge group.
In this action $\beta = \frac{1}{T}$ is the inverse 
temperature and Dirac matrices are in the representation $\gamma_{\s \mu} =
\sigma_{\s \mu}$. We have also introduced external currents coupled to the 
gauge fields. \\
We shall consider first the parity-breaking part of the fermionic part of the 
action and at this stage the scalar field is only a spectator. \\
For this purpose we will follow the approach of Fosco et al.~\cite{schap}. \\
The fermionic fields obey antiperiodic boundary conditions, while the gauge 
fields are periodic. The considered class of configurations for the gauge 
fields is:
\be
A_{\s 3}^{\s (k)} = A_{\s 3}^{\s (k)}(\tau), \quad 
A_{\s 1,2}^{\s (k)} = A_{\s 1,2}^{\s (k)} (x), \quad k=1, \ldots ,N
\ee
There is a family of gauge transformation parameters, which allow us to gauge 
the time-components  $A_{\s 3}^{\s (i)}(\tau)$ to the constants 
$a^{\s (i)}$~\cite{schap}. This makes the Dirac operator 
invariant under translations in the time coordinate (as the dependence on 
$\tau$ comes solely from the $A_{\s 3}^{\s (i)}$ fields) and therefore we 
could Fourier--expand $\psi_{\s i}$ and $\bar{\psi_{\s i}}$ over the 
Matsubara modes. Following steps, similar to those in~\cite{schap}, one finds 
that the parity-odd bit of the fermion part of the effective action is given 
by:
\ba
\Gamma_{\s odd} = \frac{i}{2\pi} \sum_{\s k, j=1}^{\s N}
\sum_{\s n=- \infty}^{\s + \infty}
\phi_{\s n}^{\s (k)} \int\epsilon_{\s lm} Q_{\s kj} 
\partial_{\s l} A_{\s m}^{\s (j)} d^{\s 2}x  
\ea
where $\phi_{\s n}^{\s (k)} = arctg\Bigl(\frac{\omega_{\s n} + Q_{\s kj}
a^{\s (j)}}{M_{\s k}}\Bigr)$ and $\omega_{\s n} = (2n + 1) 
\frac{\pi}{\beta}$ \quad is the Matsubara frequency for fermions. \\
Performing the summation we get that for $U(1) \times \ldots \times U(1)$ 
gauge group the parity-odd part of the action is:
\ba
e^{- \Gamma_{\s odd}} = \int\mathcal{D}\phi 
\ exp\Biggl\{-\frac{i}{2\pi} \sum_{\s k,j,n=1}^{\s N}
arctg\biggl[ th\Bigl(\frac{\beta M_{\s k}}{2}
\Bigr) \ tg\biggl(\frac{1}{2} \int\limits_{\s 0}^{\s \beta}
Q_{\s kj} A_{\s 3}^{\s (j)}(\tau)d \tau 
\biggr)\biggr] \nonumber \\
\times \int \epsilon_{\s lm} Q_{\s kn} \partial_{\s l} 
A_{\s m}^{\s (n)} d^{\s 2}x \nonumber \\
+ \int\limits_{\s 0}^{\s \beta} d \tau \int \Bigl[
\ (D_{\s \alpha} \phi)(D^{\s \alpha} \phi)^{*} - m^{\s 2} \phi \phi^{*} 
-  \lambda (\phi \phi^{*})^{\s 2} + j^{\s (k)}  A^{\s (k)}
\Bigr]d^{\s 2}x \Biggr\}
\ea
As the temperature $T$ approaches 0 (that is $\beta \to \infty$) this reduces
to $U(1) \times \ldots \times U(1)$ Chern--Simons gauge theory. \\
We will use now the effective parity--odd temperature dependent action 
(with the induced $U(1) \times \ldots \times U(1)$ parity breaking term)
to re-examine the confinement argument of Cornalba et al.~\cite{wil}.
First of all, let us perform the integration (using Stokes' theorem) of the 
gauge fields over the spatial co-ordinates. This gives the relation with the 
magnetic fluxes $\Phi_{\s l}$:
\ba
e^{- \Gamma_{\s odd}} = \int
&& \mkern-30mu \mathcal{D}\phi
\ exp\Biggl\{-\frac{i}{2\pi} \sum_{\s k,j,n=1 }^{\s N}
arctg\biggl[ th\Bigl(\frac{\beta M_{\s k}}{2}
\Bigr) \ tg\biggl(\frac{1}{2} \int\limits_{\s 0}^{\s \beta}
Q_{\s kj} A_{\s 3}^{\s (j)}(\tau)d \tau 
\biggr)\biggr] Q_{\s kn}\Phi_{\s n}
\nonumber \\
&& \mkern-20mu
+ \int\limits_{\s 0}^{\s \beta}\! d \tau \!\int \!\Bigl[
(D_{\s \alpha} \phi)(D^{\s \alpha} \phi)^{*} - m^{\s 2} \phi \phi^{*} 
-  \lambda (\phi \phi^{*})^{\s 2} + j^{\s (k)}  A^{\s (k)}
\Bigr]d^{\s 2}x \Biggr\}
\ea 
The equations of motion, obtained by varying the action with respect to the
magnetic fields, are:
\ba
-\frac{i}{4 \pi}
\sum_{\s k,m,n = 1}^{\s N}\frac{th\Bigl(\frac{\beta M_{\s k}}{2}\Bigr)Q_{\s kl}
Q_{\s kn}\Phi_{\s n}}
{cos^{\s 2}\biggl(\frac{1}{2}\int\limits_{\s 0}^{\s \beta} Q_{\s km}
A_{\s 3}^{\s (m)}d\tau\biggr) + th^{\s 2}\Bigl(\frac{\beta M_{\s k}}{2}\Bigr)
sin^{\s 2}\biggl(\frac{1}{2}\int\limits_{\s 0}^{\s \beta} Q_{\s km}
A_{\s 3}^{\s (m)}d\tau\biggr)} \nonumber \\
- \ q_{\s l} \int(\phi D_{\s 3} \phi^{*} - \phi^{*} D_{\s 3} \phi)
d^{\s 2}x
\ + \ \int\! \rho^{\s (l)} d^{\s 2}x =  \int\partial_{\s j}
F^{\s (l) \hskip2pt j3} \hskip2pt
d^{\s 2}x
\ea
where $\rho^{\s (l)} = j_{\s 3}^{\s (l)}$ are the charge densities. Here we 
have included explicitely the contribution of the Maxwell term 
$F_{\s \mu\nu}^{\s (i)}F^{\s (i) \hskip2pt \mu\nu}$. We ignore temperature
dependent terms which come from $O(A^{\s 4})$ terms in the effective action.
These are of higher order $\Bigl(O(Q^{\s 4})\Bigr)$ in the fermionic chagres. 
The Coulomb charges on the r.h.s. vanish because all $U(1)$ fields are 
massive. \\
Denote by $u$ the integral over the third component of the conserved N\"other 
current:  
$u = \int(\phi D_{\s 3} \phi^{*} - \phi^{*} D_{\s 3} \phi)d^{\s 2}x$
and by $C^{\s (l)}=\int \rho^{\s (l)} d^{\s 2}x$ the total external charge. 
So we have:
\be 
\mu \Phi =  C - u q 
\ee
where $\Phi=\left(\matrix{\Phi_{\s 1} \cr \vdots \cr \Phi_{\s N}}\right), \quad
q=\left(\matrix{q_{\s 1} \cr \vdots \cr q_{\s N}}\right), \quad
C=\left(\matrix{C^{\s (1)} \cr \vdots \cr C^{\s (N)}}\right)$, \quad and:
\be
\mu_{\s ln}=
\frac{i}{4 \pi}
\sum_{\s k,m = 1}^{\s N}\frac{th\Bigl(\frac{\beta M_{\s k}}{2}\Bigr)Q_{\s kl}
Q_{\s kn}}
{cos^{\s 2}\biggl(\frac{1}{2}\int\limits_{\s 0}^{\s \beta} Q_{\s km}
A_{\s 3}^{\s (m)}d\tau\biggr) + th^{\s 2}\Bigl(\frac{\beta M_{\s k}}{2}\Bigr)
sin^{\s 2}\biggl(\frac{1}{2}\int\limits_{\s 0}^{\s \beta} Q_{\s km}
A_{\s 3}^{\s (m)}d\tau\biggr)} 
\ee
As in~\cite{wil} there is another condition which must be satisfied by the 
magnetic fluxes. The Higgs field $\phi$ should be completely condensed,
i.e. $\phi(x) = v e^{\s i\sigma(x)}$, where $\sigma(x)$ is the Goldstone
boson field (the mass and the coupling constants of the
scalar field are temperature--dependent). In order that this holds we have to 
require that the covariant derivative of the scalar field vanishes. After 
integration we get:
\be
2 \pi l = q_{\s 1} \Phi_{\s 1} + \ldots + q_{\s N} \Phi_{\s N} 
= \phantom{q}^{\s t}q \ \Phi
\ee
where $2 \pi l$ is the non-trivial holonomy of the Goldstone boson (reflecting 
a topological property of the Higgs field).
Combining the two conditions (7) and (9) for the fluxes we get:
\ba
\mu \Phi & = & C - u q \nonumber \\
2 \pi l & = & \!\!\phantom{q}^{\s t}q \ \Phi
\ea
Following the analysis of~\cite{wil} we identify $u$ as a continuous 
parameter, representing the ability of the condensate to screen the electric 
charge. \\
The matrix $\mu$ can be written as 
$\mu = \!\!\!\phantom{Q}^{\s t}Q F(\beta) Q$, where $F(\beta)$ is a diagonal
matrix with entries:
\be
F_{kj}(\beta)= 
\frac{i}{4 \pi}
\frac{th\Bigl(\frac{\beta M_{\s k}}{2}\Bigr)}
{cos^{\s 2}\biggl(\frac{1}{2}\int\limits_{\s 0}^{\s \beta} Q_{\s km}
A_{\s 3}^{\s (m)}d\tau\biggr) + th^{\s 2}\Bigl(\frac{\beta M_{\s k}}{2}\Bigr)
sin^{\s 2}\biggl(\frac{1}{2}\int\limits_{\s 0}^{\s \beta} Q_{\s km}
A_{\s 3}^{\s (m)}d\tau\biggr)} \delta_{\s kj}
\ee
As $F(\beta)$ is diagonal we can always write $\mu$ in the form:
\be
\mu = \!\!\!\phantom{Q}^{\s t}Q(T) \ Q(T)
\ee
where $Q_{\s mn}(T) = F_{\s mj}^{\s 1/2}(\beta)Q_{\s jn}$. \\
Let us now try to eliminate the screening in (10) by inverting the matrix 
$\mu$. We get that if the determinant of $\mu$ is not zero and if 
\be
\phantom{q}^{\s t}q \ \mu^{\s -1} q = 0
\ee
then the screening would be eliminated (the condition 
$\phantom{q}^{\s t}q \ \mu^{\s -1} q = 0$ is the condition for confinement,
proposed by Cornalba et al.~\cite{wil}. According to their analysis, if
the determinant of $\mu$ vanishes, then $\mu^{\s -1}$ should be interpreted
as the transposed matrix of co-factors). \\
Assuming that the determinant of $\mu$ is not zero, we can re-write this as:
\be
\phantom{Q}^{\s t}\Bigl(\tilde{Q}(T)q\Bigr)\tilde{Q}(T)q=0
\ee
where $\tilde{Q}(T)$ is the matrix of co-factors. This equation shows that the
vector $\tilde{Q}(T)q$ is orthogonal to itself (``orthogonal'' with respect 
to the matrix multiplication of column vectors) and, therefore, this is the 
null vector:
\be
\tilde{Q}(T)q=0
\ee
This is an equation for the values of the boson field charges, which would 
eliminate the screening mechanism. As we see, we can have a non--trivial 
solution if, and only if, $det \ Q = 0$, which contradicts to our initial 
assumption ($det \ \mu \ne 0$). Therefore, we cannot eliminate the screening.
Otherwise, this theory would be inconsistent with the induced parity--breaking
term. This argument is valid for all values of the temperature. \\ 
Intuitively, one can expect that if condition (9) is violated, after 
elimination of screening, there would be currents which would not fall off 
faster than $1/r$ at infinity and the resulting long--range forces will lead 
to diverging energies.
We argue that condition (9) can never be violated  --- this condition 
represents the fact that we are left with a residual symmetry after the 
spontaneous symmetry breakdown. If this condition does not hold, it would mean
that the symmetry is restored. This, on its turn, will lead to diverging 
energy straight away, but not in the broken Higgs phase.\\
We conclude that confinement is not possible in the Higgs phase in the 
presence of the dynamically generated parity--breaking term (which coincides 
with Chern--Simons term in zero--temperature limit). If there are 
configurations with infinte energy, they must necessarily be outside the 
broken Higgs phase --- where the gauge symmetry is restored.

\end{document}